\documentclass[12pt,letterpaper]{article}
\usepackage{physics}
\usepackage{amssymb}
\usepackage{authblk}
\usepackage{xcolor}
\usepackage{datetime}
\usepackage{parskip}
\usepackage{graphicx}
\usepackage[left=1in,right=1in,top=1in,bottom=1in]{geometry}

\usepackage{amsthm}
\theoremstyle{remark}
\newtheorem{rem}{Remark}

\numberwithin{equation}{section}

\mmddyyyydate

\DeclareMathOperator{\SO}{SO}

\DeclareMathOperator{\AdS}{AdS}

\title{Perturbations to Generalized Kink-like Topological Defects in $\AdS$}
\author{Orlando Alvarez\footnote{oalvarez@miami.edu}\,}
\author{Matthew Haddad\footnote{m.haddad@miami.edu}}
\affil{Department of Physics, University of Miami, 1320 Campo Sano Ave, Coral Gables, FL 33146}
\date{Last Typeset: \today~at~\currenttime}

\begin{document}
	\maketitle
	
	\begin{abstract}
	We explore perturbations to a kink-like (codimension 1) topological defect whose world brane is $\AdS_{q}$ embedded into $\AdS_{q+1}$. Previously, we found solutions in the limit the mass of the scalar field vanishes. In this article we extend a calculation previously done in $\AdS_{2}$ to higher-dimensional embedding spaces and find that all perturbations to the mass of the field are stable to first order as expected in a theory with topological defects. We find that the equation of motion to the correction strongly resembles a problem well-known in quantum mechanics.
	\end{abstract}

	\section{Introduction}
	This is the third of a series of articles in which the authors explore the dynamics of topological defects in anti de Sitter space. These objects are interesting to study as a possible consequence of symmetry breaking in the early universe leading to cosmological-scale objects such as cosmic strings~\cite{Hindmarsh1995}. In more recent years, some study has gone into how these objects might influence the dynamics of nearby massive objects~\cite{SantanaMota2019} Our model concerns an $\SO(l)$-gauged Higgs-type field theory embedded into $\AdS_{q+l}$ with $q,l \geq 1$. For convenience, we will often refer to these with the shorthand ``$(q,l)$ defect'' for the codimension-$l$ topological defect with a worldbrane of $\AdS_{q}$ embedded in $\AdS_{q+l}$. In the first article~\cite{Alvarez2018a}, we laid out a framework for finding exact analytical spherically-symmetric topological defect solutions. The method followed in the spirit of Bogomolny, Prasad, and Sommerfield (BPS)~\cite{Prasad1975,Bogomolny1976}, who were able to study exact solutions to the 'tHooft-Polyakov Monopole and Julia-Zee dyon in flat spacetime. The solutions we found were affectionately dubbed the ``double BPS'' solutions for that reason. In the double BPS solution, the finite radius of curvature of $\AdS$ sets the length-scale for the physics, allowing one to take the limit in the equations of motion that the mass of the fields fall to zero. Lugo et al.~\cite{Lugo1999,Lugo2000} and Ivanova et al.~\cite{Ivanova2008,Ivanova2017} also did studies of topological defects in $\AdS$, especially that of monopoles in the former and pure Yang Mills solutions in the latter. For a discussion on how these works relate to this model and more detail on the process, we refer the reader to the original article~\cite{Alvarez2018a}.
	
	The second article~\cite{Alvarez2018} concerned an extension of this using perturbation theory. Given a double BPS solution, we allow a small perturbation in the potential to reintroduce the mass of the field. We took the case of the (1,1) kink-like defect in particular; that is, the codimension-1 topological defect with a worldbrane that is $\AdS_{1}$ embedded in $\AdS_{2}$. We found that these defects are stable under this perturbation and we were able to calculate the correction to the defect energy that results.
	
	In this paper, we extend our discussion to a more general $(q,1)$ defect. We are able to show that these solutions are linearly stable in general as expected in a theory containing topological defects.
	
	\section{The metric and full EOM}
	
	The $(q,1)$ kink is a codimension-1 topological defect extending
	from a totally geodesic  $\AdS_{q}$ embedded in
	$\AdS_{q+1}$.  There is still only a scalar field
	$\Phi$, so we do not have a gauge field to worry about.  We take
	the metric of $\AdS_{q}$ to be the standard maximally invariant Lorentzian
	metric\footnote{We use the mostly $+$ convention for the signature.} on $\AdS_{q}$, $\dd{s}^{2}_{\AdS_{q}} =
	\gamma_{ab}(\sigma)\dd{\sigma^{a}} \otimes \dd{\sigma^{b}}$ with
	$\vb*{\sigma}$ as coordinates along the submanifold.  If $k<0$ is the
	sectional curvature of $\AdS_{q+1}$ (and by extension, the totally geodesic $\AdS_{q}$), then the radius of curvature is
	defined by $\rho= \lvert k \rvert^{-1/2}>0$.  A point in $\AdS_{q+1}$
	will be given coordinates $(\vb*{\sigma},\nu)$ where $\nu$ is the signed
	distance along a geodesic that is normal to $\AdS_{q}$.  Note 
	that $\nu \in (-\infty,+\infty)$. The
	metric in $\AdS_{q+1}$ is then
	\begin{equation}
	\dd{s}^{2}_{\AdS_{q+1}} = \cosh[2](\frac{\nu}{\rho}) 
	\dd{s}^{2}_{\AdS_{q}} + \dd{\nu}^{2}. \label{eq:fullMetric}
	\end{equation}
	
	From this point forward we will use dimensionless coordinates. These can be obtained by scaling everything by the radius of curvature: $\nu \to \rho\nu$, $\vb*{\sigma} \to \rho\vb*{\sigma}$, and $\Phi \to \Phi_{0}\,\Phi$ where $\Phi_{0}$ the vacuum expectation value for the scalar field $\Phi(\vb*{\sigma},\nu)$.
	
	The action for a $(q,1)$-defect is
	\begin{equation}
	I = -\Phi_{0}^{2} \int \dd[n]{x} \sqrt{-g} \qty(\frac{1}{2}g^{\mu\nu}\partial_{\mu}\Phi(x)\partial_{\nu}\Phi(x) + \epsilon U(\Phi(x))). \label{eq:fullAction}
	\end{equation}
	where $\epsilon$ is a bookkeeping parameter that will be used to keep track of the
	order of the perturbative expansion. We assume the potential $U(\Phi)$ is of the symmetry breaking type with minimum at $\pm 1$ after the field rescaling.
	
	This gives rise to the equation of motion:
	\begin{equation}
	\frac{1}{\sqrt{-g}}\partial_{\mu}\qty(\sqrt{-g}\,g^{\mu\nu} \partial_{\nu}\Phi) - \epsilon U'(\Phi) = 0. \label{eq:fullEOM}
	\end{equation}
	The first term is the negative of the d'Alembertian $\Box_{\AdS_{q+1}}$
	and which in  our chosen coordinates may be expressed as :
	\begin{align}
	\frac{1}{\sqrt{-g}} \partial_{\mu} \qty(\sqrt{-g} g^{\mu\nu} \partial_{\nu}) &= \frac{1}{\cosh[q](\nu)\sqrt{-\gamma}}\biggl[\partial_{a}\qty(\cosh[q](\nu)\sqrt{-\gamma} \frac{1}{\cosh[2](\nu)}\gamma^{ab}\partial_{b}) \nonumber \\
	&\quad  + \partial_{\nu}\qty(\cosh[q](\nu)\sqrt{-\gamma}\partial_{\nu})\biggr]  \nonumber \\
	&= \frac{1}{\cosh[2](\nu)} \qty(-\Box_{\AdS_{q}}) + \frac{1}{\cosh[q](\nu)}\partial_{\nu}\qty(\cosh[q](\nu)\partial_{\nu}) \nonumber \\
	\Box_{\AdS_{n}} &= \frac{1}{\cosh[2](\nu)}\Box_{\AdS_{q}} 
	- \frac{1}{\cosh[q](\nu)} \partial_{\nu} \qty(\cosh[q](\nu)\partial_{\nu}) 
	\label{eq:adsDAlembertian}
	\end{align}	
	Using this operator, equation~\eqref{eq:fullEOM} becomes:
	\begin{equation}
	\Box_{\AdS_{q+1}}\Phi = -\epsilon U'(\Phi). \label{eq:fullEOMDAlembertian}
	\end{equation}
	This is the wave equation with a source.  In the limit
	where $\epsilon\,\downarrow 0$, we obtain the wave equation
	$\Box_{\AdS_{q+1}}\Phi = 0$. In addition,  we retain 
	the boundary conditions 
	\begin{equation}
		\Phi \xrightarrow{\nu\to \pm\infty} \pm 1
		\label{eq:dBPS-BC}
	\end{equation}
	that guarantee a kink-like solution.
	We seek solutions
	with localized energy, that is, energy density that falls off exponentially as
	$\nu\to\pm\infty$.  It does not need to be localized in the
	$\vb*{\sigma}$ direction(s). 
	
	We found maximally symmetric solutions that depend only on the
	normal coordinate $\nu$, and we denoted them by $\phi(\nu)$. These are the so-called double BPS solutions and they satisfy the wave equation
	\begin{equation}
	\Box_{\AdS_{q+1}}\phi(\nu) = 0, \label{eq:zeroOrderEOM}
	\end{equation}
	with boundary conditions $\phi(\nu) \to \pm 1$ as $\nu \to
	\pm\infty$, and is discussed in \cite{Alvarez2018a}.  We are
	interested in analyzing the corrections to our solutions by
	introducing a perturbation about the double BPS solution, and we write the field
	as the double BPS solution and a perturbation: $\Phi = \phi +
	\epsilon \psi$.
	
	The first-order correction $\psi(\vb*{\sigma},\nu)$ obeys a wave equation 
	with a source term determined by the zeroth order double BPS 
	solution:
	\begin{equation}
	\Box_{\AdS_{q+1}}\psi = -U'(\phi). \label{eq:firstOrderEOM}
	\end{equation}
	The general solution will consist of a particular solution to
	the inhomogeneous equation and the general solution to the
	homogeneous equation: $\psi = \xi(\nu) + \eta(\vb*{\sigma},\nu)$, where $\xi(\nu)$ solves the
	inhomogeneous equation with a source and $\eta(\vb*{\sigma},\nu)$ solves the
	homogeneous equation.  Our ansatz for the perturbative solution is
	therefore $\Phi = \phi + \epsilon\qty(\xi + \eta)$.
	
	This was only to first order, but it's not too hard to generalize
	this procedure to any order in perturbation theory.  We can show
	that, in general:
	\begin{equation}
	\Box_{\AdS_{q+1}}\Phi_\qty[j+1] = -\epsilon U'\qty(\Phi_{\qty[j]}) +
	\order{\epsilon^{j+2}}, \label{eq:iterativeEOM}
	\end{equation}
	where we use $\Phi_{[j]}$ to denote the $j$th order solution to
	the equations of motion.  This gives an iterative scheme to
	solving the differential equation.  At each order you will have to
	specify exactly how to choose the solution to the inhomogeneous
	equation because the d'Alembertian operator has a non-trivial
	kernel.  To uniquely specify a perturbative solution we use the
	same method as in the $q=1$ case.
		
	\section{Solving the EOM}
	\subsection{Separating the EOM}
	The first order contribution $\psi(\vb*{\sigma},\nu)$ must satisfy
	\eqref{eq:firstOrderEOM}.  Since the source term on the right is
	only a function of $\nu$, we can write $\psi(\vb*{\sigma},\nu) =
	\xi(\nu) + \eta(\vb*{\sigma},\nu)$, where $\xi(\nu)$ is a particular
	solution to the inhomogeneous equation and
	$\eta(\vb*{\sigma},\nu)$ is a solution to the homogeneous equation below:
	\begin{equation}
	\frac{1}{\cosh[q](\nu)} \dv{\nu}(\cosh[q](\nu)\dv{\xi}{\nu}) =
	U'(\phi),
	\label{eq:diffeqXi}
	\end{equation}
	\begin{equation} 
	\frac{1}{\cosh[2](\nu)}\Box_{\AdS_{q}} \eta - \frac{1}{\cosh[q](\nu)}\pdv{\nu}(\cosh[q](\nu)\pdv{\eta}{\nu}) = 0. \label{eq:diffeqEta}
	\end{equation}
	The boundary conditions are $\xi(\nu)
	\xrightarrow{\abs{\nu}\to\infty}0$ and $\eta(\vb*{\sigma},\nu)
	\xrightarrow{\abs{\nu} \to \infty}0$.
	Equation~\eqref{eq:diffeqXi} is an ordinary linear differential
	equation with source and we immediately discuss its solution.
	Afterwards, we will discuss how to solve PDE \eqref{eq:diffeqEta},
	
	\subsection{Solving for $\xi(\nu)$}
	
	The solution to \eqref{eq:diffeqXi} depends upon the choice of
	potential $U(\phi)$.  The solution in the case of the $q=1$
	defect in the symmetry-breaking potential $U(\phi) =
	\frac{\Lambda}{8}(\phi^{2} - \phi_{0}^{2})^{2}$ was explored in
	our previous article~\cite{Alvarez2018}.  We restate the results
	here, and extend the discussion of this toy model to a few other
	cases.  The procedure for solving the equation is nearly identical
	for $q > 1$, though the difficulty in obtaining the solution
	varies significantly as $q$ increases, see
	Appendix~\ref{sec:solvingInhom} for details.
	
	In the case of $q=1$, the ODE becomes:
	\begin{equation}
	\xi''(\nu) + \tanh(\nu) \xi'(\nu) -
	\frac{\tilde{\mu}^{2}}{\pi^{3}} \arctan(\sinh(\nu))
	\qty(4\arctan(\sinh(\nu))^{2} - \pi^{2}) = 0.
	\label{eq:odeXi_1-1}
	\end{equation}
	Here, $\tilde{\mu}^{2} = \Lambda\phi_{0}^{2}\rho^{2}$ is the
	squared flat-space mass of the scalar field $\phi$ scaled by the
	radius of curvature in order to keep dimensionless coordinates.
	The solution to this equation with the given boundary conditions
	is
	\begin{multline}
	\xi_{1}(\nu) = \frac{\tilde{\mu}^{2}}{5 \pi ^3} \left(-20 \text{gd}^3(\nu ) \log \biggl(1+e^{2 i \text{gd}(\nu )}\right)-5 i \pi ^2 \text{gd}^2(\nu ) \\
	-5 i \left(\pi ^2-12 \text{gd}^2(\nu )\right) \text{Li}_2\left(e^{4 i \tan ^{-1}\left(e^{\nu }\right)}\right) +5 \pi ^2 \text{gd}(\nu ) \left(\log \left(\frac{\cosh (\nu )}{2}\right)+2 \log \left(1+e^{2 i \text{gd}(\nu )}\right)\right) \\
	-90 \text{gd}(\nu ) \text{Li}_3\left(e^{4 i \tan ^{-1}\left(e^{\nu }\right)}\right)+90 \zeta (3) \text{gd}(\nu )  -60 i \text{Li}_4\left(e^{4 i \tan ^{-1}\left(e^{\nu }\right)}\right)-i \pi ^4\biggr), \label{eq:solnXi_1-1}
	\end{multline}
	where $\text{Li}_{n}(x)$ is the polylogarithm function of order
	$n$, and $\text{gd}$ is the Gudermannian function.

\begin{figure}[tbp]
		\centering
		\includegraphics[width=0.7\linewidth]{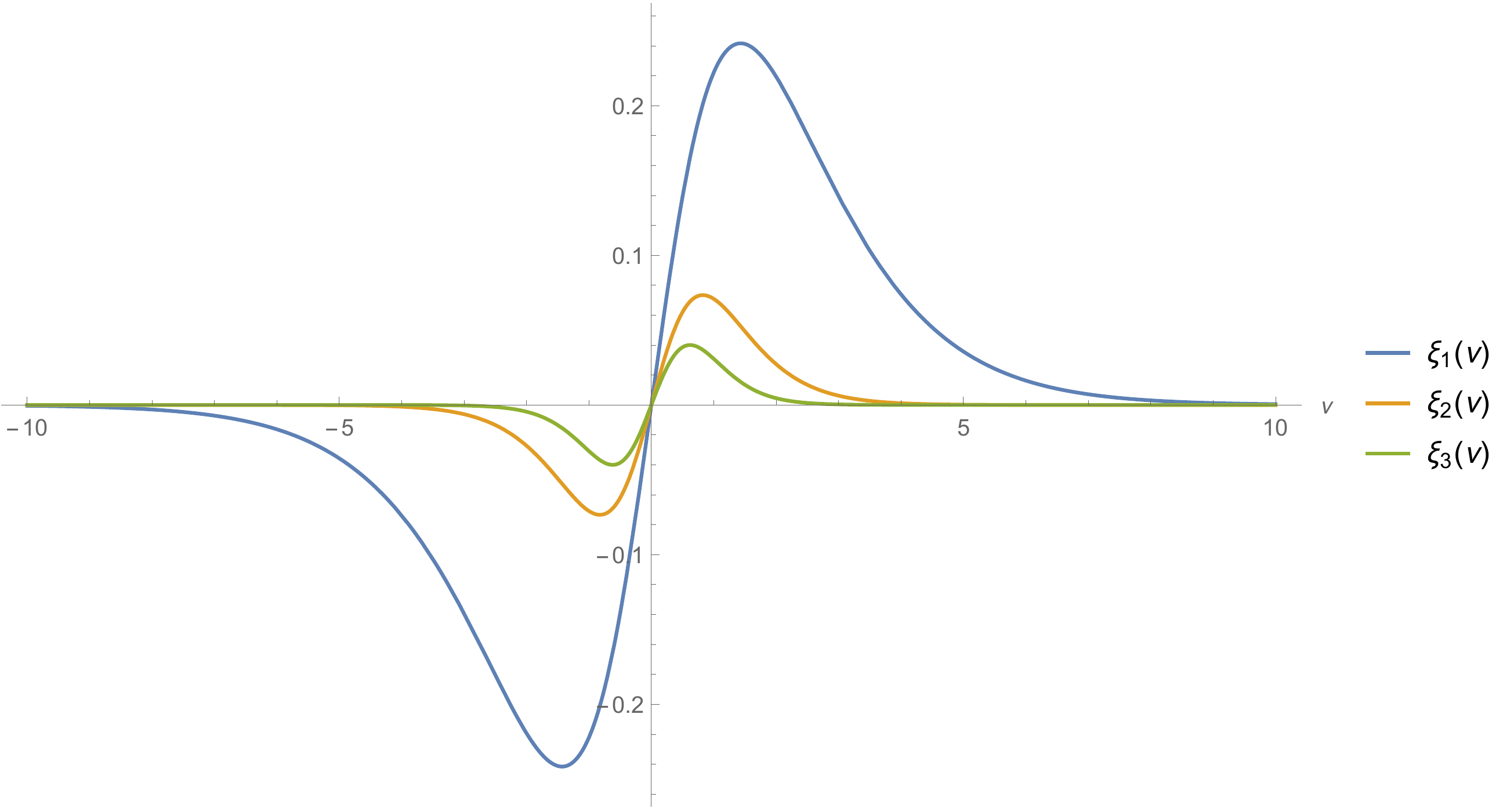}
		\caption{Plot of the real valued functions $\xi_{q}(\nu)$ for
		$\tilde{\mu}=1$ and $q=1,2,3$.  }
		\label{fig:xi-123-1-correction-nu}
	\end{figure}

	When $q=2$ the ODE is given in \eqref{eq:odeXi_2-1}.  The solution
	to this equation with the appropriate boundary conditions is
	\begin{equation}
	\xi_{2}(\nu) = \frac{\tilde{\mu}^{2}}{4} \left(2 \nu -\tanh (\nu ) \log \left(4 \cosh ^2(\nu )\right)\right) = \frac{\tilde{\mu}^{2}}{2} \qty(\nu - \tanh(\nu)\log(2\cosh(\nu))). \label{eq:solnXi_2-1}
	\end{equation}
	Though not immediately obvious, $\xi_{1}(\nu)$ is real-valued, see
	Appendix~\ref{sec:solvingInhom}.  We also obtained an explicit
	expression for the case $q=3$.  This expression is lengthy and
	does not appear to yield novel information.  It is included for
	completeness in the aforementioned appendix.  These solutions are
	plotted in figure~\ref{fig:xi-123-1-correction-nu}.
	
	\subsection{Solving the equation for $\eta(\vb*{\sigma},\nu)$}
	
	We focus on solving the PDE~\eqref{eq:diffeqEta} for $\eta$.  We
	use the separation of variables method and try a separable
	solution $\eta(\vb*{\sigma},\nu) = u(\vb*{\sigma})w(\nu)$.
	Substituting into the equation, we obtain
	\begin{equation}
	\frac{\Box_{\AdS_{q}}u}{u} - \frac{1}{w\cosh[q-2](\nu)}\dv{\nu}(\cosh[q](\nu)\dv{w}{\nu}) = 0. \label{eq:diffeqEtaSeparated}
	\end{equation}
	
	The first term involves only $\vb*{\sigma}$-derivatives and the second only involves $\nu$-derivatives. Since these are independent, the only way they can cancel exactly is if they are equal to some constant $\lambda$. The first term gives:
	\begin{equation}
	\Box_{\AdS_{q}}u = -\lambda u. \label{eq:diffeqU}
	\end{equation}
	
	The second term gives:
	\begin{equation}
	-\frac{1}{w} \frac{1}{\cosh[q-2](\nu)} \dv{\nu}(\cosh[q](\nu) \dv{w}{\nu}) = \lambda,
	\end{equation}
	which can be rewritten as
	\begin{equation}
	\dv[2]{w}{\nu} + q\tanh(\nu)\dv{w}{\nu} = -\lambda \sech[2](\nu)w \label{eq:diffeqW}
	\end{equation}
	This is a one-dimensional problem.  There is a gauge
	transformation we can do to eliminate the first derivative term
	and turn this equation into a P\"{o}schl-Teller model in 
	Schr\"odinger form.
	
	\subsection{The P\"{o}schl-Teller model and the equation for
	$w(\nu)$} 
	
	Consider the first order differential operator
	\begin{equation}
	D = \dv{\nu} + \frac{q}{2}\tanh(\nu). \label{eq:opD}
	\end{equation}
	Squaring this operator leads to
	\begin{equation}
	D^{2} = \dv[2]{\nu} + q \tanh(\nu) \dv{\nu} +
	\qty(\frac{q}{2}\sech[2](\nu) + \frac{q^{2}}{4}\tanh[2](\nu)).
	\label{eq:opDSquare}
	\end{equation}
	
	We can see that the first two terms are exactly the operator we
	have in \eqref{eq:diffeqW}, and with a little bit of algebra and 
	hyperbolic function identities we can rewrite \eqref{eq:diffeqW} as
	\begin{equation}
	-D^{2}w - \qty(\lambda + \frac{q^{2}}{4}-
	\frac{q}{2})\sech[2](\nu)w = -\frac{q^{2}}{4}w.
	\label{eq:diffeqWDsub}
	\end{equation}
	
	The covariant derivative like operator $D$ can be turned into an
	ordinary derivative via a gauge transformation.  We note that
	\begin{equation}
	D = \cosh[-q/2](\nu) \dv{\nu} \cosh[q/2](\nu), 
	\label{eq:opDRewrite}
	\end{equation}
	and
	\begin{equation}
	D^{2}w = \cosh[-q/2](\nu)\dv[2]{\nu}(\cosh[q/2](\nu)w). \label{eq:opDSquareRewrite}
	\end{equation}
	
	If we define the auxiliary function $\widetilde{w}(\nu) = \cosh[q/2](\nu)w(\nu)$ then we have
	\begin{equation}
	D^{2}w = \sech[q/2](\nu) \dv[2]{\widetilde{w}}{\nu}. \label{eq:opDAuxW}
	\end{equation}	
	Substituting \eqref{eq:opDAuxW} into \eqref{eq:diffeqWDsub} gives
	\begin{equation} -\dv[2]{\widetilde{w}}{\nu} - \qty(\lambda +
	\frac{q^{2}}{4} - \frac{q}{2})\sech[2](\nu)\widetilde{w} =
	-\frac{q^{2}}{4}\widetilde{w}.  \label{eq:diffeqWPT}
	\end{equation}
	
	This is a Schr\"odinger-type differential equation with a P\"oschl-Teller-style potential 
	\begin{equation}
	V(\nu) = -\qty(\lambda+\frac{q(q-2)}{4})\sech[2](\nu), \label{eq:PTPotential}
	\end{equation}
	with eigenvalue of $-q^{2}/4$.  The strategy is to determine all
	values of the parameter $\lambda+\frac{q(q-2)}{4}$ that lead to an
	eigenvalue $-q^{2}/4$.  We work out the solution to this problem
	in Appendix~\ref{sec:1DPT}.
	For brevity, the normalized solution found there is:
	\begin{align}
	w_{m}(\nu) &= 
	\frac{\qty(-1)^{m}}{\mathcal{N}_{q/2}\;  \sqrt{(m+1)!\,\qty(q+2)^{\overline{m}}}} 
	\cosh[m+1](\nu)\qty(\sech(\nu)\dv{\nu})^{m}\sech[q+1](\nu), 
		\label{eq:solnW} \\
	\mathcal{N}_{q/2}^{2} &= \int_{-\infty}^{\infty} \dd{\nu} \sech[q](\nu), 
	\nonumber
	\end{align}
	where $m= 0,1,2,3,4,\dotsc$, and $x^{\overline{m}} =
	x(x+1)(x+2)\cdots(x+m-1)$ is the rising factorial of $x$.  The
	values of $\lambda$ that lead to an  eigenvalue of $-q^{2}/4$ for 
	\eqref{eq:diffeqWPT} are
	the infinite sequence 
	\begin{equation}
		\lambda_{m} = m(m+q-1) \text{ with $m=0,1,2,3,\dotsc$.}
		\label{eq:lambda-m}
	\end{equation}
	We note that a consequence of $q \ge 2$ and $m \ge 0$ is that $\lambda_{m} \ge 0$.
	
	\subsection{The Eigenvalue Problem for $u(\vb*{\sigma})$}
	
	Next we study equation~\eqref{eq:diffeqU} incorporating 
	the correct expression for the eigenvalues  \eqref{eq:lambda-m}.
	The equation for $u(\vb*{\sigma})$ 
	to be solved is
	\begin{equation}
	\qty\Big(\Box_{\AdS_{q}} + \lambda_{m})u = 0, 
	\qquad \lambda_{m} = m(m+q-1). \label{eq:diffeqUKG}
	\end{equation}

	This is the massive Klein-Gordon equation on the submanifold
	$\AdS_{q}$ with mass  $\sqrt{\lambda_{m}}$ that depends
	on the dimension $q$ and the index $m$.
	We can work out the explicit form of $\Box_{\AdS_{q}}$. First, write the metric for $\AdS_{q}$ as
	\begin{equation}
	\dd{s}_{\AdS_{q}}^{2} = -\cosh[2](r) \dd{t}^{2} + \dd{r}^{2} + \sinh[2](r) \dd{s}_{S^{q-2}}^{2}. \label{eq:metricAdSq}
	\end{equation}
	
	In these coordinates, we have split $\vb*{\sigma}$ into the
	coordinates $t$, $r$ and $(q-2)$ angles $\vb*{\theta}$ on
	$S^{q-2}$.  The metric determinant is therefore
	\begin{equation}
	\sqrt{-\det \gamma} = \cosh(r)\sinh[q-2](r), 
	\label{eq:metricDetAdSq}
	\end{equation}
	and \eqref{eq:diffeqUKG} becomes
	\begin{equation}
	\qty[-\frac{1}{\cosh[2](r)} \pdv[2]{t} + \qty(\tanh(r) +
	(q-2)\coth(r)) \pdv{r} + \pdv[2]{r} + \frac{1}{\sinh[2](r)}
	\laplacian_{S^{q-2}} - \lambda_{m}] u = 0.
	\label{eq:diffeqULaplacian}.
	\end{equation}
	The metric is diagonal, so we try  separation of variables with ansatz
	\begin{equation}
	u(\vb*{\sigma}) = R_{mJ\omega}(r)Y^{J}_{M}(\vb*{\theta})
	e^{-i\omega t}\,,
	\label{eq:ansatzU}
	\end{equation}
	where $R_{mJ\omega}(r)$ is the radial wave function, and $Y^{J}_{M}$ are
	orthonormal real spherical harmonics on $S^{q-2}$.  Here $J = 0,
	1,2,3,\dotsb$ is the angular momentum that labels the irreducible
	representation of $\SO(q-1)$.  The index $M$ takes
	\begin{equation*}
		(2J+q-3) \, \frac{(J+q-4)!}{J! (q-3)!}\;.
	\end{equation*}
	different values.  Note that for $q=3$, the formula above gives
	$2$, which is the dimensionality of the real irreducible
	representations of $\SO(2)$.
	
	For now, we will suppress the indices.  If we plug this back in
	and divide the whole equation by $RYe^{-i\omega t}$ per the usual
	procedure in separation of variables, we  wind up with the
	equation:
	\begin{equation}
	-\dv[2]{R}{r} - [\tanh(r) + (q-2)\coth(r)]\dv{R}{r} - \frac{\omega^{2}R}{\sinh[2](r)} + \frac{\ell R}{\cosh[2](r)} + \lambda R = 0, \label{eq:diffeqR}
	\end{equation}
	where $\ell = J(J+q-3)$ is an eigenvalue of the Laplacian on 
	$S^{q-2}$. Note that the ODE is  parametrized by $q$, $J$, $m$, 
	$\omega$. We ignore $q$ because it is fixed, and this is why we 
	had previously written $R_{mJ\omega}$.
	
	We have managed to reduce our problem to an ODE in $r$. Since this is a 1D problem, let's try a gauge transformation (as we did prior) to see if we can eliminate the first derivative terms:
	\begin{align}
	D &\equiv \dv{r} + \frac{1}{2}[\tanh(r) + (q-2)\coth(r)], \label{eq:opDNew} \\
	D^{2} &= \dv[2]{r} + \qty(\tanh(r) + (q-2)\coth(r))\dv{r} +	\left(\frac{1}{2}\sech[2](r) - \frac{q-2}{2}\csch[2](r) + \frac{1}{4}\tanh[2](r) \right. \nonumber \\
	&\quad \left. + \frac{\qty(q-2)^{2}}{4}\coth[2](r) + \frac{q-2}{2}\right). \label{eq:opDSquaredNew}
	\end{align}

	The term inside the large parentheses is
	\begin{multline}
	\frac{1}{2}\sech[2](r) - \frac{q-2}{2}\csch[2](r) + 
	\frac{1}{4}\tanh[2](r) + \frac{\qty(q-2)^{2}}{4}\coth[2](r) + 
	\frac{q-2}{2} \\ = \frac{1}{4}\sech[2](r) + 
	\frac{1}{4}\qty[\qty(q-2)^{2}-2(q-2)]\csch[2](r) + 
	\frac{1}{4}\qty(q-1)^{2}
	\end{multline}

	Putting all this together, we have
	\begin{multline}
	\dv[2]{r} + \qty(\tanh(r) + (q-2)\coth(r))\dv{r} = \\
	D^{2} - \qty[\frac{1}{4}\sech[2](r) + \frac{(q-2)(q-4)}{4}\csch[2](r) + 
	\frac{\qty(q-1)^{2}}{4}] \label{eq:diffeqRDSub}
	\end{multline}
	We can use this to rewrite \eqref{eq:diffeqR} as
	\begin{multline}
	-D^{2}R + \qty(\frac{1}{4} - \omega^{2})\sech[2](r) R \\
	+ \qty(\frac{1}{4}(q-2)(q-4) + \ell)\csch[2](r) R +  \qty(\frac{1}{4}\qty(q-1)^{2} + \lambda) R = 0 \label{eq:diffeqRPartialGauge}
	\end{multline}
	To eliminate the covariant derivative, let $\Upsilon =
	e^{\frac{1}{2}\ln(\cosh(r))}e^{\frac{q-2}{2}\ln(\sinh(r))}$, then
	$D = \Upsilon^{-1} \dv{r} \Upsilon$.  If we also set
	$\widetilde{R}(r) = \Upsilon R(r)$ 
	then we can rewrite
	\eqref{eq:diffeqRPartialGauge} as
	\begin{multline}
	-\dv[2]{\widetilde{R}}{r} - \qty(\omega^{2} - \frac{1}{4})\sech[2](r) 
	\widetilde{R} \\
	+ \qty(\ell + \frac{1}{4}(q-2)(q-4))\csch[2](r) 
	\widetilde{R} + \qty(\lambda + \frac{1}{4}\qty(q-1)^{2})\widetilde{R} 
	= 0 \label{eq:our-gen-PT-x}
	\end{multline}
	It should be mentioned that the Hilbert space that 
	$\widetilde{R}$ belongs to is plain old square integrable 
	functions on $(0,\infty)$. The reason is that $\Upsilon(r)^{2}$ 
	is just the volume element factor \eqref{eq:metricDetAdSq} that 
	enters the inner product required for the function $R$.
	We next substitute the basic parameters into the above and 
	algebraically simplify to obtain
	\begin{multline}
	-\dv[2]{\widetilde{R}}{r} - \qty(\omega^{2} - \frac{1}{4})\sech[2](r) 
	\widetilde{R} \\
	+ \qty(\qty(J + \frac{q-3}{2})^{2} - \frac{1}{4}) \csch[2](r) 
	\widetilde{R} = - \qty(m+ \frac{q-1}{2})^{2} \widetilde{R} 
	\label{eq:our-gen-PT}
	\end{multline}
	This is a Schr\"odinger-type equation with a generalized
	P\"oschl-Teller type potential.  We examine how to analyze the
	spectrum for this type of problem for $q \ge 3$ in
	Appendix~\ref{sec:NDPT}.  The question we address may be phrased
	as follows. 
		\begin{rem}
			\label{rem:params} \sffamily We assume that the immutable
			parameter $q$ is \emph{a priori} fixed by the
			dimensionality of space-time.  There are two parameters at
			our disposal in \eqref{eq:our-gen-PT}: the parameter $m$
			is fixed by the solution of a previous eigenvalue problem
			\eqref{eq:lambda-m}, and the parameter $J$ is fixed by the
			rotational properties of the solution.  How do you find
			$\omega$ and $\widetilde{R}$ to make equation
			\eqref{eq:our-gen-PT} valid?
		\end{rem}
	You can recast the question in more familiar language by observing
	that if you multiply both sides of eq.~\eqref{eq:our-gen-PT} by
	$\cosh^{2} r$, then the above is an eigenvalue problem for a
	complicated second order ODE with regular singular points
	depending on parameters $m$ and $J$ with eigenvalue $\omega^{2} -
	1/4$ and eigenfunction $\widetilde{R}$ to be determined.
	
	The case $q=2$ is special and does not require new tools because $J=0$ automatically, consequently the
	``centrifugal barrier potential term'' $1/\sinh^{2} r$ is absent.  The hamiltonian for the radial wave function
	$\widetilde{R}$ is just a P\"{o}schl-Teller hamiltonian of the
	type discussed in Appendix~\ref{sec:1DPT}.  You can analyze this
	problem in two equivalent ways: (1) Take the range of $\nu$ to be
	$(-\infty,+\infty)$.  The hamiltonian is parity invariant so the
	energy eigenfunctions may be taken to also be eigenfunctions of
	parity.  (2) Mimic the discussion for $q \ge 3$, by taking the
	range of $\nu$ to be $[0,\infty)$ and thinking of the symmetry
	group of $S^{q-2} =\{+1,-1\}$ as being the group with two elements
	$\{+1,-1\}$.  The role of the spherical harmonics can be replaced
	studying two classes of functions.  Those that vanish at $r=0$ and
	functions whose derivative vanishes at $r=0$.  In viewpoint (1),
	there are respectively the even functions and the odd functions.
		
	\section{Conclusion}
	We have found the first order correction to the double BPS
	solution $\phi(\nu)$ of a generalized $(q,1)$ kink-like
	topological defect under a small perturbation in the mass of the
	scalar field.  Putting everything together, the solution to the
	equation of motion \eqref{eq:fullEOMDAlembertian} is a linear 
	superposition
	\begin{multline}
		\Phi(t,r,\vb*{\theta},\nu) = \phi(\nu) \\
		+ \epsilon \qty(\xi(\nu) +
		\sum_{m=0}^{\infty} \sum_{\pm}\sum_{(J,\omega)\in \mathcal{A}_{m}}
		C_{mJ\omega,\pm}\, w_{m}(\nu) R_{mJ\omega}(r)\,
		Y^{J}_{M}(\vb*{\theta})\, e^{\mp i\omega\, t})+
		\order{\epsilon^{2}}, 
	\label{eq:fullSoln}
	\end{multline}
	where $\vb*{\sigma} = (t,r,\vb*{\theta})$ are the coordinates of
	$\AdS_{q}$, and where the functions $\xi(\nu)$, $R_{mJ\omega}(r)$ and
	$w_{m}(\nu)$ in the expression above take specific forms once $q$
	is specified.  The summation set $\mathcal{A}_{m}$ is described in Figure~\ref{fig:angle}. The takeaway here is that  the boundary
	conditions, which serve as a check on topological constraints,
	ensure that the function $\xi(\nu)$ die off as $\nu \to \infty$,
	the stability of these solutions depends on what values the
	parameter $\omega$ can have.  By determining that this frequency
	is bounded from below and is always a positive integer, we have
	shown that all solutions are stable to first order.  Furthermore,
	once these functions are determined, the correction to the energy
	can be calculated.  The correction to the energy was explicitly computed in the case of the
	$(1,1)$ defect in our previous article~\cite{Alvarez2018}.
	
	We mention an interesting topic that we have not explored. The case with $m=0$ corresponds to $\lambda_{0}=0$ in the Klein-Gordon eq.~\eqref{eq:diffeqUKG}. Looking at the full solution \eqref{eq:fullSoln}, we see that this corresponds to a linearized perturbation of the form $w_{0}(\nu) R_{0J\omega}(r)Y^{J}_{M}(\vb*{\theta})\, e^{\mp i\omega\, t}$.  This excitation is bound to $\AdS_{q}$ because the factor $w_{0}(\nu)$ decays exponentially in the direction normal to $\AdS_{q}$ with the length scale set by the radius of curvature. This excitation has a factor  $R_{0J\omega}(r)Y^{J}_{M}(\vb*{\theta})\, e^{\mp i\omega\, t}$ which corresponds to a massless particle on $\AdS_{q}$. From the viewpoint of $\AdS_{q}$ this looks like a massless excitation, but from the bulk view of $\AdS_{q+1}$ these are massive excitations.
	
	\appendix
	\section{Solving the P\"{o}schl-Teller equation} \label{sec:PT}
	
	\subsection{1D problem \label{sec:1DPT}}
	Begin by defining ladder-like operators for this problem:
	\begin{equation}
	\mathsf{a}_{\beta} = \dv{x} + \beta \tanh(x), \qquad \mathsf{a}^{\dagger}_{\beta} = -\dv{x} + \beta \tanh(x) \label{eq:def-ladder}
	\end{equation}
	We then compute:
	\begin{equation}
			\mathsf{a}^{\dagger}_{\beta} \mathsf{a}_{\beta} =
			-\dv[2]{x} -\beta(\beta+1)\sech[2](x)+\beta^{2}
			\label{eq:adagger-a}
	\end{equation}
	
	The P\"{o}schl-Teller hamiltonian operator is $\mathsf{h}_{\beta}
	= -\dv[2]{x} - \beta(\beta+1)\sech[2](x)$.  In our problem we need
	eigenfunctions with a negative eigenvalue, thus we require
	$-\beta(\beta + 1) < 0$.  The hamiltonian is invariant under
	$\beta \to -(\beta+1)$, so we restrict to the parameter range to
	$\beta>0$.  We are interested in the eigenvalue problem
	$\mathsf{h}_{\beta} \ket{\beta,\epsilon} = \epsilon
	\ket{\beta,\epsilon}$.  We note that
	\begin{equation}
		\mathsf{a}^{\dagger}_{\beta}\mathsf{a}_{\beta} =
		\mathsf{h}_{\beta} + \beta^{2}
		\label{eq:adag-a}
	\end{equation}
	Two immediate consequences of these equations is that 
	\begin{equation}
		\mathsf{a}_{\beta+1}\mathsf{a}^{\dagger}_{\beta+1} =
		\mathsf{h}_{\beta} + (\beta+1)^{2}
	\label{eq:a-adagger}
	\end{equation}
	Two immediate consequences of these equations are that
	$\expval{\mathsf{h}_{\beta}} \ge -\beta^{2}$ and
	$\expval{\mathsf{h}_{\beta}} \ge -(\beta+1)^{2}$.  Remember that
	we are in the parameter range $\beta>0$.  If $\epsilon$ is an
	eigenvalue of $\mathsf{h}_{\beta}$, then eigenvalues are forbidden
	in the region $\epsilon < -\beta^{2}$ and in the region $\epsilon
	< - (\beta+1)^{2}$.  Since the latter region is a subset of the
	first, we conclude that if $\epsilon$ is an eigenvalue of
	$\mathsf{h}_{\beta}$, then we have the rigorous Rayleigh
	eigenvalue lower bound $\epsilon \ge -\beta^{2}$.
	
	We can  verify two identities
	\begin{align}
		\mathsf{h}_{\beta-1}\bigl( \mathsf{a}_{\beta} 
		\ket{\beta,\epsilon}\bigr) &=
		\epsilon \bigl( \mathsf{a}_{\beta}\ket{\beta,\epsilon}\bigr) 
		\label{eq:PT-a}\\
		\mathsf{h}_{\beta+1}\qty(\mathsf{a}^{\dagger}_{\beta+1}
		\ket{\beta,\epsilon}) &=
		\epsilon\qty(\mathsf{a}^{\dagger}_{\beta+1}\ket{\beta,\epsilon})
		\label{eq:PT-adag}
	\end{align}
	We see that $\mathsf{a}_{\beta}\ket{\beta,\epsilon}$ is an
	eigenstate of $\mathsf{h}_{\beta-1}$ with the same energy
	eigenvalue $\epsilon$.  Also $\mathsf{a}^{\dagger}_{\beta+1}
	\ket{\beta,\epsilon}$ is an eigenstate of $\mathsf{h}_{\beta+1}$ 
	with the same eigenvalue $\epsilon$.
	
\begin{figure}[tbp]
	\centering
	\includegraphics[width=0.7\textwidth]{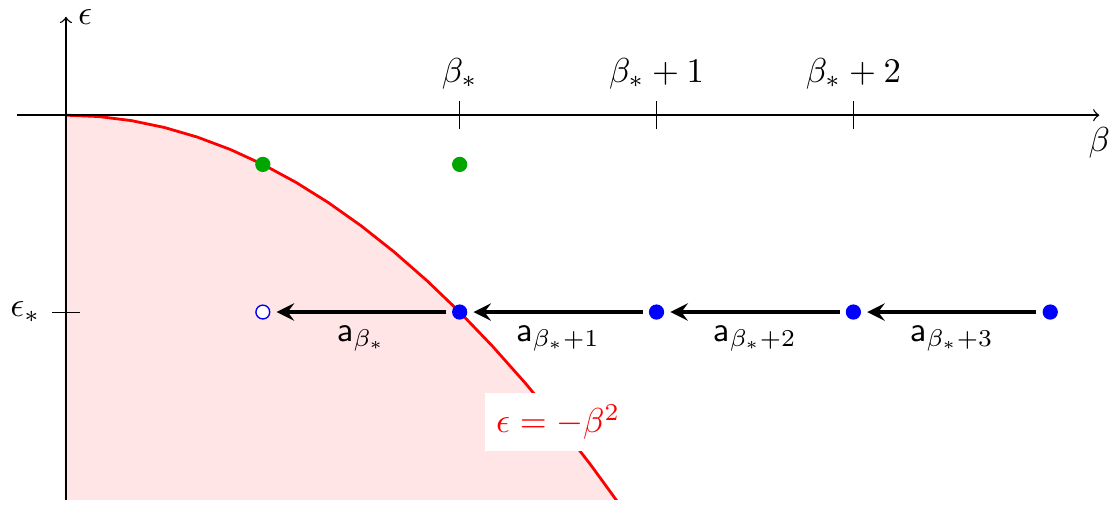}
	\caption{In the coupling-eigenvalue $(\beta,\epsilon)$-plane for
	the P\"{o}schl-Teller model, the pink region $\epsilon <
	-\beta^{2}$ is forbidden by the rigorous Rayleigh eigenvalue lower
	bound.  At the point $(\beta,\epsilon)$, the lowering ladder
	operator $\mathsf{a}_{\beta}$ moves you horizontally to the left
	while the raising ladder operator $\mathsf{a}_{\beta+1}^{\dagger}$
	moves you horizontally to the right according to equations
	\eqref{eq:PT-a} and \eqref{eq:PT-adag}.  It is a consequence of
	the analysis discussed in the text that $\beta_{*}$ is on the
	boundary curve.}
	\label{fig:PT-hopping}
\end{figure}
		
Assume we have found an eigenvector
$\ket{\beta_{\sharp},\epsilon_{*}}$ of the hamiltonian
$\mathsf{h}_{\beta_{\sharp}}$.  The state $$\mathsf{a}_{\beta_{\sharp}-k+1}\dotsb \mathsf{a}_{\beta_{\sharp}-2}
\mathsf{a}_{\beta_{\sharp}-1} \mathsf{a}_{\beta_{\sharp}}
\ket{\beta_{\sharp},\epsilon_{*}}$$ is in principle an eigenvector of
$\mathsf{h}_{\beta_{\sharp}-k}$ with the same eigenvalue
$\epsilon_{*}$.  The big issue is that for $k$ sufficiently large you
will cross into the forbidden pink region of
figure~\ref{fig:PT-hopping} and the state construction process has to
stop.  In other words, there exists an eigenvector
$\ket{\beta_{*},\epsilon_{*}} \neq 0$ with the point
$(\beta_{*},\epsilon_{*})$ in the allowed region, but
$\mathsf{a}_{\beta_{*}} \ket{\beta_{*},\epsilon_{*}}=0$ because the
point $(\beta_{*}-1,\epsilon_{*})$ is in the forbidden region.  Note
that as a consequence of \eqref{eq:adag-a}, we have that
$\mathsf{h}_{\beta_{*}} \ket{\beta_{*},\epsilon_{*}} = -\beta_{*}^{2}
\ket{\beta_{*},\epsilon_{*}}$ and we conclude that $\epsilon_{*}=
-\beta_{*}^{2}$.  In other words, the point $(\beta_{*},\epsilon_{*})$
must be on the boundary of the Rayleigh bound given by the red
curve $\epsilon=-\beta^{2}$.  To find all values of $\beta$ with the
same energy $\epsilon_{*}$, the procedure is to start at the
Rayleigh bound by choosing $\beta_{*} = \sqrt{-\epsilon_{*}}$ and
operate on the state $\ket{\beta_{*},\epsilon_{*}}$ with a product of
appropriately indexed $\mathsf{a}_{\bullet}^{\dag}$ that move you
horizontally to the right, and obtain the infinite sequence
$\beta_{*},\beta_{*}+1,\beta_{*}+2, \dotsc$.  \emph{The problem of
finding all the eigenvectors and eigenvalues of
$\mathsf{h}_{\beta_{*}}$ is a different problem, and the solution is
indicated without explanation by asking at how you would construct the
green circles in the figure.  For the illustrated value of
$\beta_{*}=2$, there are only two eigenvectors of $\mathsf{h}_{2}$
with eigenvalues $-1$ and $-4$.}

	We now determine an explicit formula for the state
	$\ket{\beta_{*},\epsilon_{*}}$.  Let $\varphi_{\beta_{*}}(x) =
	\braket{x}{\beta_{*},\epsilon}$ then $\mathsf{a}_{\beta_{*}} 
	\varphi_{\beta_{*}}(x)=0$ implies
	\begin{equation*}
		\qty(\dv{x} + \beta_{*}\tanh(x))\varphi_{\beta_{*}} = 0
	\end{equation*}
	The normalized solution is easily found to be
	\begin{equation}
	\varphi_{\beta_{*}}(x) = \mathcal{N}^{-1}_{\beta_{*}} \sech[\beta_{*}](x), 
	\text{ where } \mathcal{N}_{\beta_{*}}^{2} = 
	\int_{\infty}^{\infty} \dd{x} \sech[2\beta_{*}](x) 
	\label{eq:beta-ground-fn-normal} 
	\end{equation}
	Now we will construct all the other choices of $\beta$ that have
	the same $\epsilon=\epsilon_{*}= -\beta_{*}^{2}$.  Consider a
	normalized state $\ket{\beta,\epsilon_{*}}$.  We can apply
	$\mathsf{a}^{\dagger}_{\beta+1}$ to this state and get the
	relation $\mathsf{a}_{\beta+1}^{\dagger}\ket{\beta,\epsilon_{*}} =
	\mathcal{N}_{\beta+1}\ket{\beta+1,\epsilon_{*}}$, where the 
	normalization constant is determined via
	\begin{align}
	\abs{\mathcal{N}_{\beta+1}}^{2}
	\braket{\beta+1,\epsilon_{*}}{\beta+1,\epsilon_{*}} &=
	\bra{\beta+1,\epsilon_{*}}\mathsf{a}_{\beta+1}\mathsf{a}^{\dagger}_{\beta+1}
	\ket{\beta+1,\epsilon_{*}} 
	\nonumber \\
	\abs{\mathcal{N}_{\beta+1}}^{2} &=
	\bra{\beta+1,\epsilon_{*}}\left(\mathsf{h}_{\beta} +
	\qty(\beta+1)^{2} \right) \ket{\beta+1,\epsilon_{*}} 
	 \nonumber \\
	&= \mel{\beta+1,\epsilon_{*}}{\mathsf{h}_{\beta}}{\beta+1,\epsilon_{*}}
	+ \qty(\beta+1)^{2} 
	\nonumber \\
	&=\epsilon_{*} + \qty(\beta+1)^{2}  \nonumber \\
	&= -\beta_{*}^{2} + (\beta+1)^{2}  \nonumber \\
	\abs{\mathcal{N}_{\beta+1}}^{2} &= 
	\qty(\beta + 1 -\beta_{*})\qty(\beta + 1 + \beta_{*}) 
	\label{eq:normal-beta+1}
	\end{align}
	We can generalize equation \eqref{eq:normal-beta+1} to any
	$\beta=\beta_{*}+m$, with $m = 0,1,2,3,\dotsc$:
	\begin{equation}
	\abs{\mathcal{N}_{\beta_{*}+m}}^{2} = \qty(m+1) \qty(2\beta_{*} + m + 1) \label{eq:normal-beta+m}
	\end{equation}
	With this factor, we can now write down an ``excited'' state as
	\begin{align}
	\ket{\beta_{*}+m,\epsilon} &= \frac{1}{\mathcal{N}_{\beta_{*}+m}} \mathsf{a}^{\dagger}_{\beta_{*}+m} \ket{\beta_{*} + m - 1,\epsilon} \nonumber \\
	&= \frac{1}{\mathcal{N}_{\beta_{*}+m}\mathcal{N}_{\beta_{*}+m-1}} \mathsf{a}^{\dagger}_{\beta_{*}+m}\mathsf{a}^{\dagger}_{\beta_{*}+m-1} \ket{\beta_{*}+m-2,\epsilon} \nonumber \\
	\ket{\beta_{*}+m,\epsilon} &= \frac{1}{\mathcal{N}_{\beta_{*}+m}\mathcal{N}_{\beta_{*}+m-1}\cdots\mathcal{N}_{\beta_{*}+1}} \mathsf{a}^{\dagger}_{\beta_{*}+m} \mathsf{a}^{\dagger}_{\beta_{*}+m-1}\cdots \mathsf{a}^{\dagger}_{\beta_{*}+1} \ket{\beta_{*},\epsilon} \label{eq:general-beta-ket}
	\end{align}
	We can obtain explicit expressions for the eigenfunctions. We need to work out
	$\braket{x}{\beta_{*}+m,\epsilon}$.  We note that
	\begin{equation}
	\mathsf{a}^{\dagger}_{\beta} = -\cosh[\beta](x) \dv{x} \cosh[-\beta](x) 
	\label{eq:a-dagger-cosh}
	\end{equation}
	We need to figure out what the product of all the operators
	amounts to.  For example,
	\begin{align}
	\mathsf{a}_{\beta_{*}+3}^{\dagger} \mathsf{a}^{\dagger}_{\beta_{*}+2} \mathsf{a}^{\dagger}_{\beta_{*}+1} &= \qty(-1)^{3} \cosh[\beta_{*}+3](x) \dv{x} \cosh[-\beta_{*}-3](x)\cosh[\beta_{*}+2](x) \nonumber \\
	&\quad \times \dv{x} \cosh[-\beta_{*}-2](x)\cosh[\beta_{*}+1](x)\dv{x} \cosh[-\beta_{*}-1](x) \nonumber \\
	&= \qty(-1)^{3} \cosh[\beta_{*}+3](x) \dv{x} \sech(x) \dv{x} \sech(x) \dv{x} \sech[\beta_{*}+1](x) \nonumber
	\end{align}
	Generalizing the observation above, we can write the function:
	\begin{align}
	\braket{x}{\beta_{*}+m,\epsilon} &= \left(\prod_{l=1}^{m}
	\frac{1}{\mathcal{N}_{\beta_{*}+l}} \right) \qty(-1)^{m}
	\cosh[\beta_{*}+m+1](x)
	\qty(\sech(x)\dv{x})^{m}\sech[\beta_{*}+1](x)
	\mathcal{N}_{\beta_{*}}^{-1}\sech[\beta_{*}](x) \nonumber \\
	\varphi_{\beta_{*}+m}(x) &= \left(\prod_{l=0}^{m}
	\frac{1}{\mathcal{N}_{\beta_{*}+l}} \right) \qty(-1)^{m} \cosh[\beta_{*}+m+1](x)
	\qty(\sech(x)\dv{x})^{m}\sech[2\beta_{*}+1](x)
	\label{eq:beta+m-function}
	\end{align}
	We have to make a note that $\mathcal{N}_{\beta_{*}}$ is computed
	differently than the other normalization constants.
	Next we observe that in the product
	\begin{equation}
	\mathcal{N}_{\beta_{*}+1}^{2}\mathcal{N}_{\beta_{*}+2}^{2}\cdots\mathcal{N}_{\beta_{*}+m}^{2}
	= \qty[2(2\beta_{*}+2)]
	\qty[3(2\beta_{*}+3)]\cdots\qty[(m+1)(2\beta_{*}+m+1)] 
	\end{equation}
	there are factors that combine
	\begin{align*}
	\qty(m+1)! &= (m+1)(m)(m-1)\cdots 2 \\
	\qty(2\beta_{*}+2)^{\overline{m}} &=
	(2\beta_{*}+m+1)(2\beta_{*}+m)\cdots(2\beta_{*}+2) \\
	\end{align*}
	Thus,
	\begin{equation}
	\mathcal{N}_{\beta_{*}+1}^{2}\mathcal{N}_{\beta_{*}+2}^{2}\cdots\mathcal{N}_{\beta_{*}+m}^{2} = (m+1)! \qty(2\beta_{*}+2)^{\overline{m}} \label{eq:normal-factorial}
	\end{equation}
	So if we put this back in, our normalized eigenfunctions are
	\begin{align}
	\varphi_{\beta_{*}+m}(x) &=
	\frac{\qty(-1)^{m}}{\mathcal{N}_{\beta_{*}}\;\sqrt{(m+1)!\,
	\qty(2\beta_{*}+2)^{\overline{m}}}}
	\cosh[\beta_{*}+m+1](x)\qty(\sech(x)\dv{x})^{m}\sech[2\beta_{*}+1](x)
	\label{eq:eigenfunction-PT-full}
	\end{align}

	In summary, for a fixed eigenvalue $\epsilon_{*} = -\beta_{*}^{2}$,
	we have found all possible values of $\beta$ where the associated 
	P\"{o}schl-Teller model has an eigenfunction with the same 
	eigenvalue
	\begin{equation}
	\qty(-\dv[2]{x} - \beta(\beta-1)\sech[2](x))\varphi_{\beta}(x) =
	-\epsilon_{*} \varphi_{\beta}(x)\,.
	\end{equation}
	Our solution was that $\beta = \beta_{*} + m$, where $m
	=0,1,2,3,\dotsc$.  Thus, the equation takes the form
	\begin{align}
	\qty(-\dv[2]{x} - (\beta_{*}+m)(\beta_{*}+m-1) \sech[2](x)) 
	\varphi_{\beta_{*}+m}(x) = -\beta_{*}^{2} \varphi_{\beta_{*}+m}(x)
	\end{align}
	Remember that $\varphi_{\beta_{*}+m}(x) = 
	\braket{x}{\beta_{*}+m,\epsilon_{*}}$.

	\subsection{Making contact with equation~\eqref{eq:diffeqWPT}}
	
	We can make contact with our original equation \eqref{eq:diffeqWPT} by noting that
	\begin{align}
	(\beta_{*}+m)(\beta_{*}+m-1) = \lambda_{m} + \frac{q^{2}}{4} - 
	\frac{q}{2}, \nonumber
	\intertext{therefore,}
	\qty(\beta_{*} + m -\frac{q}{2})\qty(\beta_{*} + m + \frac{q}{2} -
	1) &= \lambda_{m} \label{eq:lambda-spectrum-expression}
	\end{align}
	In our problem, the energy eigenvalue $\epsilon_{*} = -\beta_{*}^{2} =
	-\frac{q^{2}}{4}$ is fixed by the dimension of $\AdS_{q}$.  This
	reduces the expression above to
	\begin{equation}
	\lambda_{m} = m(m+q-1), \qquad m = 0,1,2,3,\dotsc
	\label{eq:lambda-spectrum-final}
	\end{equation}
	Since $\beta_{*}$ always has the fixed value of $q/2$, we index
	the solutions differently to simplify the notation.  The solution
	to \eqref{eq:diffeqWPT} previously written as
	$\varphi_{\beta_{*}+m}(x)$ will be simplified to
	\begin{align}
	\widetilde{w}_{m}(\nu) &= \frac{\qty(-1)^{m} }{\mathcal{N}_{q/2}\;
	\sqrt{(m+1)!\, \qty(q+2)^{\overline{m}}}}
	\cosh[\frac{q}{2}+m+1](\nu)\qty(\sech(\nu)\dv{\nu})^{m}\sech[q+1](\nu), \\
	\abs{\mathcal{N}_{q/2}}^{2} &= \int_{-\infty}^{\infty}
	\dd{\nu} \sech[q](\nu) \nonumber
	\end{align}
	Note that as $\lvert\nu\rvert \to \infty$, we have that
	$\widetilde{w}_{m}(\nu) \sim e^{-(q/2)\lvert \nu\rvert}$ as
	expected for a state with $\epsilon_{*} = -(q/2)^{2}$.  In terms of
	the original function $w$ we have
	\begin{align}
	w_{m}(\nu) &= \frac{\qty(-1)^{m} }{\mathcal{N}_{q/2}\;
	\sqrt{(m+1)!\,\qty(q+2)^{\overline{m}}}}
	\cosh[m+1](\nu)\qty(\sech(\nu)\dv{\nu})^{m}\sech[q+1](\nu),\label{eq:w-solution} \\
	\abs{\mathcal{N}_{q/2}}^{2} &= \int_{-\infty}^{\infty} \dd{\nu}
	\sech[q](\nu)\nonumber
	\end{align}
	Note that the parity of $\widetilde{w}_{m}$ or $w_{m}$ is
	$(-1)^{m}$ as expected.
	
	\section{The two parameter P\"{o}schl-Teller Model}
	\label{sec:NDPT}

	The eigenvalue problem for the two parameter P\"{o}schl-Teller 
	model is to find solutions of the ODE
	\begin{equation}
	-\dv[2]{\psi}{r} - \frac{\beta(\beta+1)}{\cosh[2](r)}\; \psi + 
	\frac{\gamma(\gamma+1)}{\sinh[2](r)} \;\psi = \epsilon\; \psi
	\label{eq:gen-PT}
	\end{equation}
	In our problem of interest we have $q \ge 3$ and therefore the
	eigenvalue in the right hand side of
	equation~\eqref{eq:our-gen-PT} satisfies $\epsilon \le -1$.
	Consequently, we require the function $\psi(r) \sim
	e^{-\sqrt{-\epsilon}\, r} \sim e^{-(m+(q-1)/2)r}$ as $r\to
	\infty$.  The Hilbert space consists of the square integrable
	function in the domain $(0,\infty)$.  Define the two parameter
	P\"{o}schl-Teller hamiltonian by
	\begin{equation}
	h_{\beta,\gamma} = -\dv[2]{r} - \beta(\beta+1) \sech[2](r) + 
	\gamma(\gamma+1)\csch[2](r) 
	\label{eq:hbg}
	\end{equation}
	The parameter space for the hamiltonian is the
	$(\beta,\gamma)$-plane.  Note that the hamiltonian is invariant
	under the two distinct transformation of parameters: (1) $\beta \to
	-(1+\beta)$, and (2) $\gamma \to -(1+\gamma)$.  Respectively, these 
	transformation correspond to reflection about the line 
	$\beta=-1/2$, and reflection about the line $\gamma=-1/2$. 
	Because of this symmetry we can restrict the parameter region to 
	semi-infinite rectangular region $\beta \ge -1/2$ and $\gamma \ge 
	-1/2$.
\begin{figure}[tbp]
	\centering
	\includegraphics[width=0.8\textwidth]{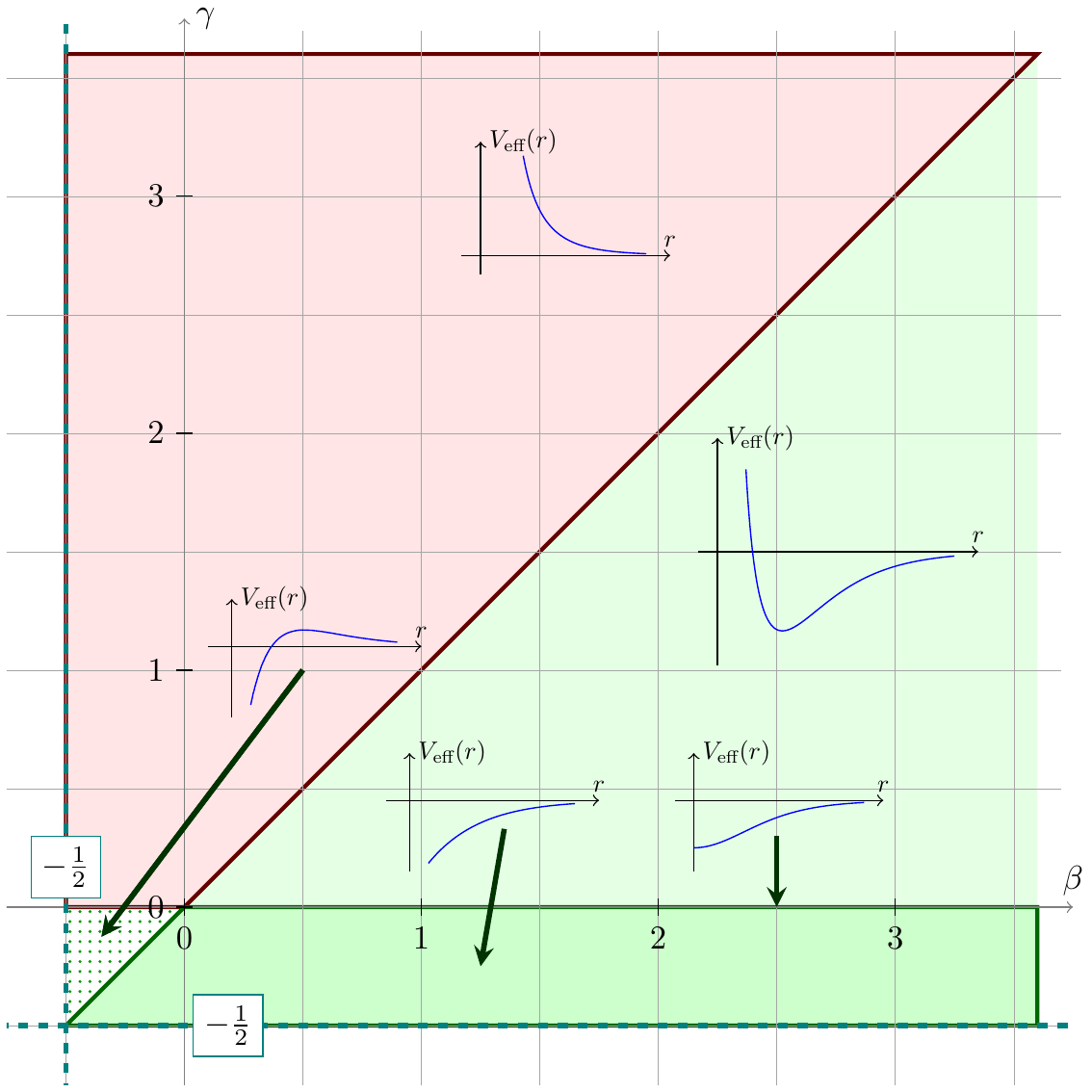}
	\caption{ Parameter space region for the hamiltonian
	$h_{\beta,\gamma}$.  The qualitative behavior of the effective
	potential $V_{\text{eff}}(r) = - \beta(\beta+1) \sech[2](r) +
	\gamma(\gamma+1)\csch[2](r)$ that enters into hamiltonian
	$h_{\beta,\gamma}$ is plotted.  The energy eigenvalue $\epsilon <
	-1$, therefore the region in pink where $V_{\text{eff}}(r)>0$ for
	all $r>0$ is ruled out because the expectation value of each
	summand in the hamiltonian $h_{\beta\gamma}$ is non-negative.  We
	expect that the values of $(\beta,\gamma)$ that admit bound states
	will be contained in the parameter regions with a green
	background.  We will pin down the region later on.}
	\label{fig:bg-parameter-space}
\end{figure}
	
	Define the operators
	\begin{align}
	a_{\beta,\gamma} &= \dv{r} + \beta\tanh(r) - (\gamma + 1)\coth(r) 
	\label{eq:def-a-bg} 
	\\
	a^{\dagger}_{\beta,\gamma} &= -\dv{r} + \beta\tanh(r) -(\gamma + 
	1)\coth(r)
	\label{eq:def-adag-bg}
	\end{align}
	
	Just as in the ordinary P\"{o}schl-Teller model we have 
	relationships
	\begin{align}
	a_{\beta,\gamma}^{\dagger} a_{\beta,\gamma} &= h_{\beta,\gamma} + 
	\qty(\beta-\gamma-1)^{2} 
	\label{eq:adag-a-bg} \\
	a_{\beta+1,\gamma-1} a_{\beta+1,\gamma-1}^{\dagger} &= h_{\beta,\gamma} 
	+ \qty(\beta-\gamma+1)^{2} 
	\label{eq:a-adag-bg}
	\end{align}
	These equations lead to Rayleigh bounds that constrain the
	eigenvalues:
	\begin{align}
		\ev{h_{\beta,\gamma}} &\geq -\qty(\beta-\gamma-1)^{2} 
		\label{eq:bound-1}\\
		\ev{h_{\beta,\gamma}} &\geq -\qty(\beta-\gamma+1)^{2}	
		\label{eq:bound-2}
	\end{align}
	
	\subsection{Relationship to our problem}
	
\begin{figure}[tbp]
			\centering
		\includegraphics[width=0.5\textwidth]{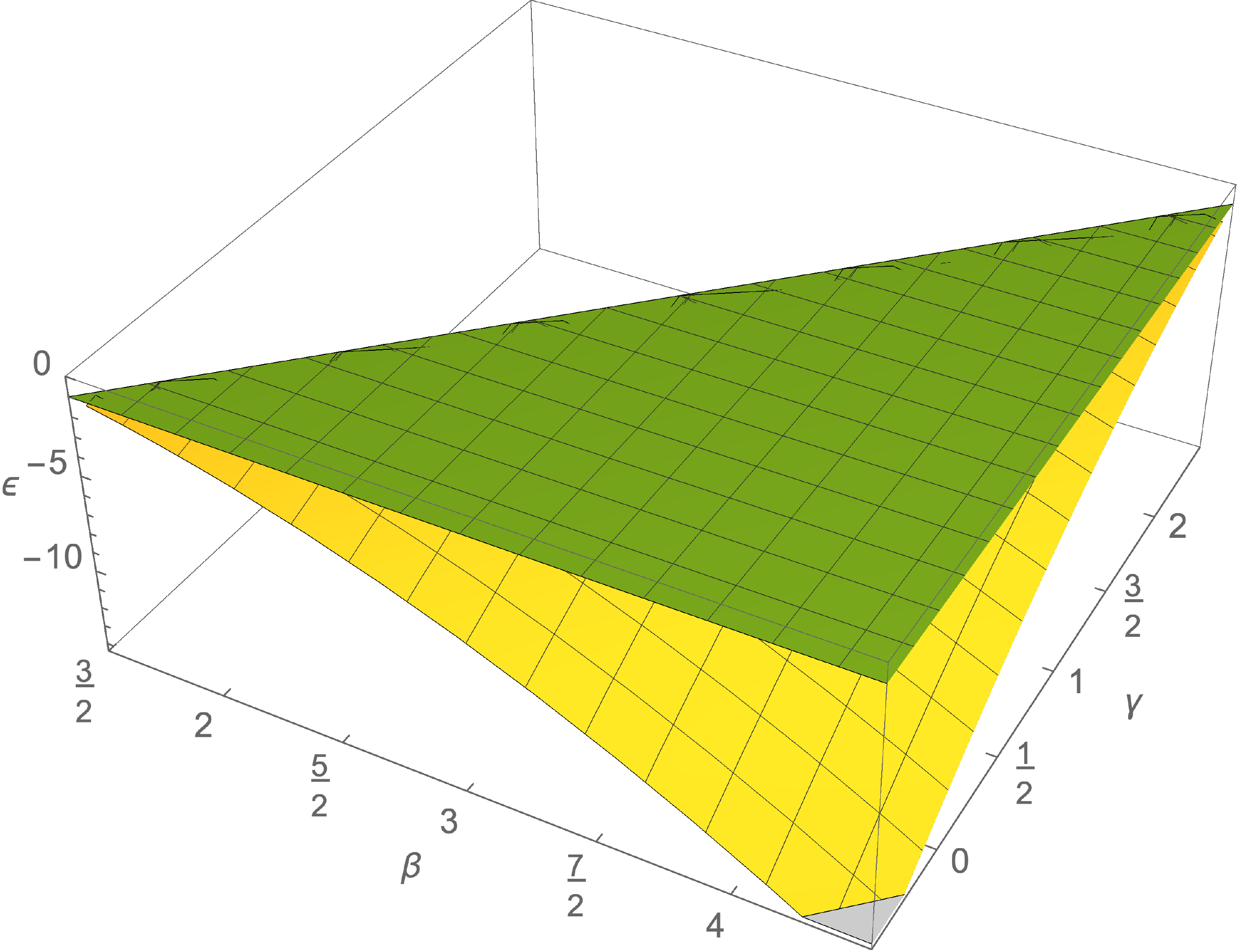}
		\caption{The wedge shaped region $\mathcal{W}$ in
		$(\beta,\gamma,\epsilon)$ parameter space is between the green
		shaded plane $\epsilon=-1$, and the yellow parabolic cylinder
		$\epsilon = -(\beta-\gamma-1)^{2}$.  It is defined by
		$\mathcal{W} = \lbrace (\beta,\gamma,\epsilon)\;:\;
		(\beta>3/2) \land (\gamma>-1/2) \land (\epsilon \le -1) \land
		(\epsilon \ge - (\beta-\gamma-1)^{2}) \rbrace$.  The symbol
		$\land$ is the \emph{``logical and''} operator.  The straight
		line that appears in the plot is the intersection of the
		horizontal plane $z=1$ with the vertical plane
		$\beta-\gamma=2$.}
		\label{fig:parameter-wedge}
	\end{figure}

	The original equation we are interested in solving is 
	\eqref{eq:our-gen-PT}. Comparing to \eqref{eq:gen-PT} we see that 
	the parameters are related by:	
		\begin{align}
			\epsilon &= - \left( m + \frac{q-1}{2}\right)^{2}\,, 
			\label{eq:eps-p}\\
			\beta(\beta+1) &= \omega^{2} - \frac{1}{4}\,, 
			\label{eq:beta-p} \\
			\gamma(\gamma+1) &= \left( J + \frac{q-3}{2}\right)^{2} - 
			\frac{1}{4} \,.		
			\label{eq:gamma-p}	
		\end{align}
	It is worthwhile recollecting that the three parameters that
	appear in \eqref{eq:our-gen-PT} are all integers, and are
	constrained by $m \ge 0$, $J\ge 0$, and $q \ge 3$.  In addition, we
	chose the parameter range of $\beta$ and $\gamma$ to be $\beta +
	1/2 \ge 0$ and $\gamma + 1/2 \ge 0$.  The first observation we
	make is that we are interested in studying generalized
	P\"{o}schl-Teller equation when the eigenvalue 
	\begin{equation}
		\epsilon \le -1\,.
		\label{eq:e-bound}
	\end{equation} 
	Next we observe that eqs. \eqref{eq:beta-p} and 
	\eqref{eq:gamma-p} may be rewritten as
		\begin{align*}
			\left(\beta + \frac{1}{2}\right)^{2} &= \omega^{2} \,, \\
			\left(\gamma + \frac{1}{2}\right)^{2} &= \left( J + 
			\frac{q-3}{2}\right)^{2} \,.
		\end{align*}
	The advantage of writing the equations in this form is that they 
	automatically incorporate the parameter range of $\beta$ and 
	$\gamma$, and the reflection symmetry about $\beta=-1/2$ and 
	$\gamma=-1/2$. When we take the square root of the two equations 
	above, the parameter range tells us that we have to take the 
	positive branch. Time reversal allows us to restrict to $\omega 
	\ge 0$. Note that $J +(q-3)/2 \ge 0$ so we have
		\begin{align*}
			\beta + \frac{1}{2} &= \omega \,, \\
			\gamma + \frac{1}{2} &= J + 
			\frac{q-3}{2} \,.
		\end{align*}
	In summary we have that
		\begin{align}
			\epsilon &= - \left( m + \frac{q-1}{2}\right)^{2}\,, 
			\label{eq:eps-p-1}	\\
			\gamma &= J + \frac{q-4}{2} \ge \frac{q-4}{2}\,, 
			\label{eq:g-J-q} \\
			\beta  &= \omega - \frac{1}{2} \ge -\frac{1}{2}\,. 
			\label{eq:b-omega} 
		\end{align}
Note that for even integer $q \ge 4$, we have that $\gamma\ge 0$ is an
integer.  For an odd integer $q \ge 3$ we have that $\gamma \ge -1/2$
and $\gamma$ is a half-integer, \emph{i.e.}, $\gamma \in \mathbb{Z} +
\frac{1}{2}$.  Figure~\ref{fig:bg-parameter-space} tells that that the
values of $(\beta,\gamma)$ that give bound states will be found in the
parameter region that is complimentary to the shaded pink region where
$V(r) >0$ for $r>0$.  A study of equations \eqref{eq:eps-p-1},
\eqref{eq:g-J-q}, \eqref{eq:b-omega}, and bounds \eqref{eq:bound-1}
and \eqref{eq:bound-2} produces a wedge shaped region $\mathcal{W}$ in
$(\beta,\gamma,\epsilon)$ parameter space, see
Figure~\ref{fig:parameter-wedge}.  Solutions to the question posed in
Remark~\ref{rem:params} must belong to $\mathcal{W}$.

\subsection{The ladder}

If $\ket{\beta,\gamma,\epsilon}$ is an eigenvector of 
$h_{\beta,\gamma}$ with eigenvalue $\epsilon$ then you can verify the 
``ladder'' relations:
\begin{align}
	h_{\beta-1,\gamma+1} \bigl( a_{\beta,\gamma} 
	\ket{\beta,\gamma,\epsilon}\bigr) = \epsilon\, \bigl( a_{\beta,\gamma} 
	\ket{\beta,\gamma,\epsilon}\bigr) 
	\label{eq:action-a}\\
	h_{\beta+1,\gamma-1} \bigl( a_{\beta+1,\gamma-1}^{\dagger}
	\ket{\beta,\gamma,\epsilon}\bigr) = \epsilon\, \bigl(
	a_{\beta+1,\gamma-1}^{\dagger} \ket{\beta,\gamma,\epsilon}\bigr)
	\label{eq:action-adag}
\end{align}
In summary, the state $a_{\beta,\gamma}
\ket{\beta,\gamma,\epsilon}\bigr)$ is an eigenvector of
$h_{\beta-1,\gamma+1}$ with the same eigenvalue $\epsilon$, and the
state $a_{\beta+1,\gamma-1}^{\dagger} \ket{\beta,\gamma,\epsilon}$ is
an eigenvector of $h_{\beta+1,\gamma-1}$ with the same eigenvalue
$\epsilon$. We also have the rigorous Rayleigh bound on the 
eigenvalues of $h_{\beta,\gamma}$:
\begin{equation}
	\epsilon \ge
	-(\beta-\gamma-1)^{2}.
	\label{eq:ray-bound}
\end{equation}

\begin{figure}[tbp]
		\centering
	\includegraphics[width=0.5\textwidth]{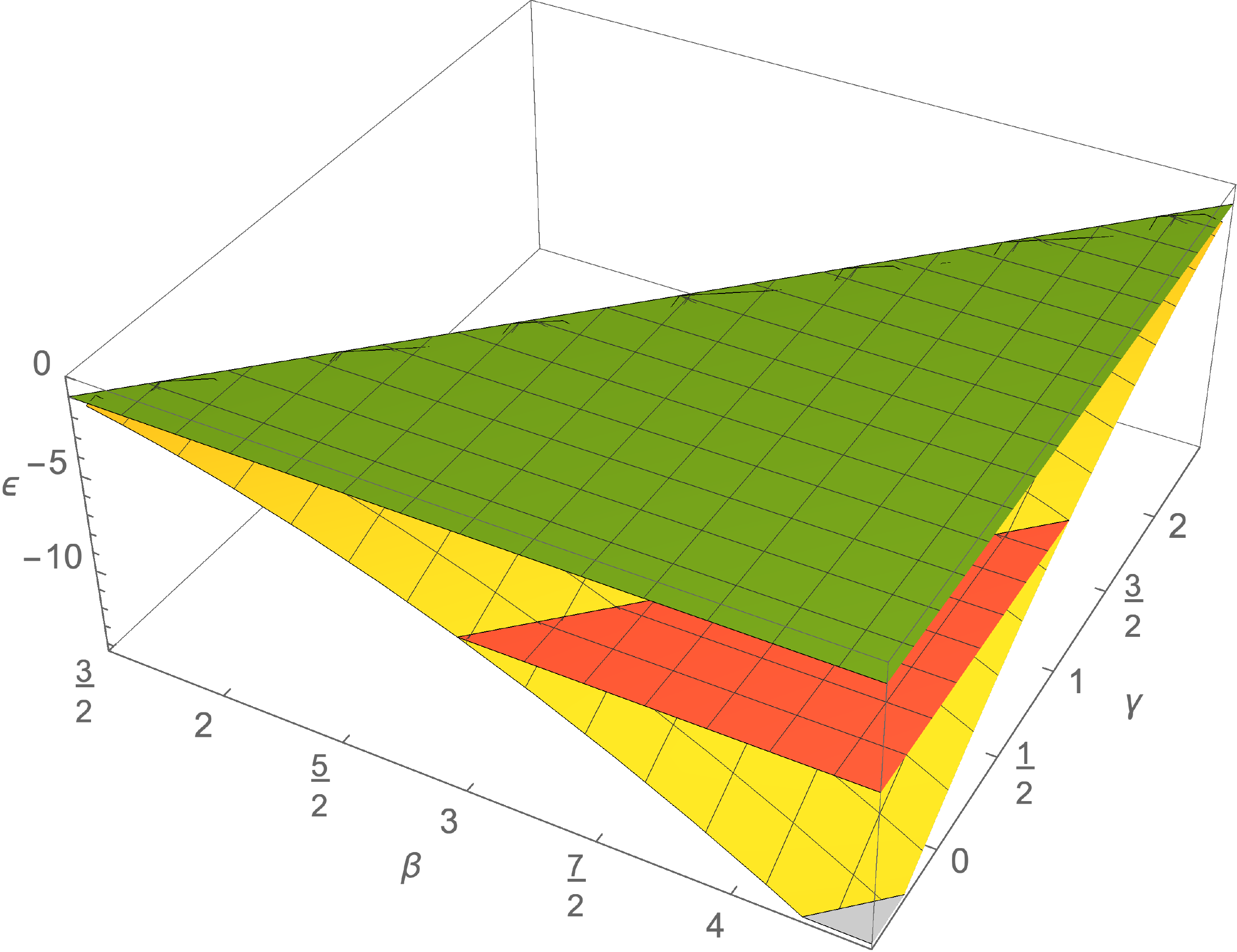}
	\caption{Once you choose a value of $m$, the energy eigenvalue
	$\epsilon$ is determined by eq.~\eqref{eq:eps-p-1}.  Let
	$\epsilon_{*}$ be one such value.  The horizontal reddish region
	is the intersection of the horizontal plane
	$\epsilon=\epsilon_{*}$ with the wedge $\mathcal{W}$. The 
	appropriate ladder operators move you diagonally in the 
	$\epsilon=\epsilon_{*}$ plane.}
	\label{fig:parameter-wedge-eps}
\end{figure}
\begin{figure}[tbp]
	\centering
	\includegraphics[width=0.7\textwidth]{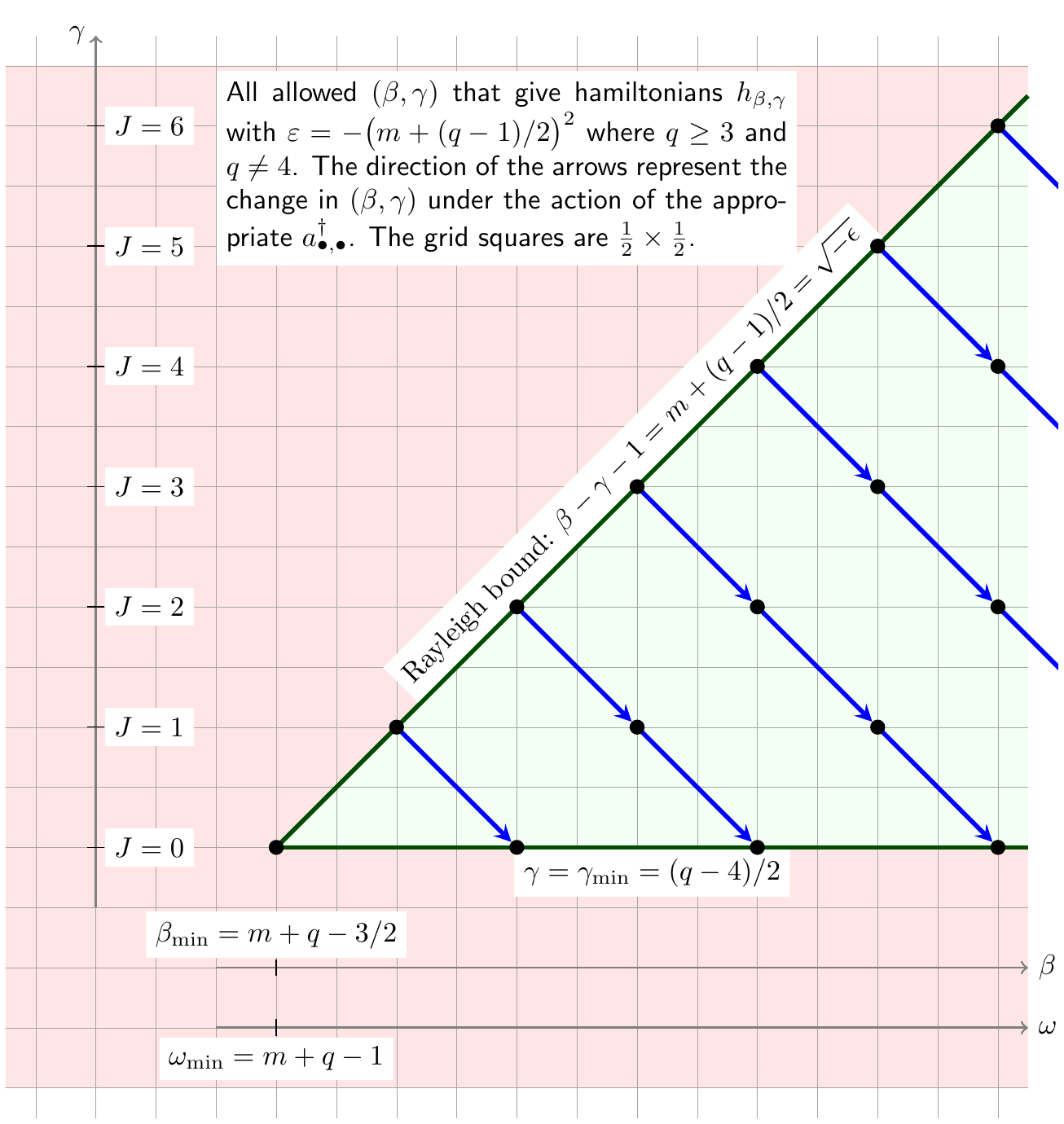}
	\caption{The angle and its interior determine the allowed values
	of $\omega$ that provide normalizable solutions to
	eq.~\eqref{eq:our-gen-PT}.  The required value of $\omega$ at each
	point is simply $\omega = \beta + 1/2$.  The $(\beta,\gamma)$
	allowed parameter region is $\lbrace (\beta,\gamma)\;:\; \gamma
	\ge (q-4)/4 \text{ and } \beta-\gamma-1 \ge m + (q-1)/2 \rbrace$.
	The vertex of the angle is at the point
	$\bigl(m+q-3/2,(q-4)/2\bigr)$.  Every circle on the Rayleigh bound
	gives a solution of type \eqref{eq:bound-sol}. The set of lattice 
	points denoted by the black circles will be denoted by 
	$\mathcal{A}_{m}$.}
	\label{fig:angle}
\end{figure}

First we observe that according to eq.~\eqref{eq:eps-p-1}, choosing
$m$ determines energy eigenvalue $\epsilon_{*}$.  Our task is to
determine all $(\beta,\gamma)$ that give eigenvectors with eigenvalue
$\epsilon_{*}$.  This means that we are interested in admissible
choices of $\beta$ and $\gamma$ that belong to the red plane in
Figure~\ref{fig:parameter-wedge-eps}, and we restrict our selves to
that plane.  We also observe that the different admissible $\gamma$
are determined by \eqref{eq:g-J-q} with $J=0,1,2,\dotsc$.  All we need
is to determine the admissible $\beta$.  Assume there is a state
$\ket{\beta_{\sharp},\gamma_{\sharp},\epsilon_{*}}$ that is an
eigenvector of $h_{\beta_{\sharp},\gamma_{\sharp}}$ with eigenvalue
$\epsilon_{*}\le -1$.  Operating on this state with a product of
appropriate $a_{\bullet,\bullet}$ moves us diagonally to the NW. In
Figure~\ref{fig:angle}, the direction of the arrows indicate how the
operators $a_{\bullet,\bullet}^{\dagger}$ change the value of
$(\beta,\gamma)$ according to eq.~\eqref{eq:action-adag}.  Here we
are interested in the action of $a_{\bullet,\bullet}$ which according
to \eqref{eq:action-a} would correspond to arrows in
Figure~\ref{fig:angle} with the opposite orientation.  Under the
action of the various $a_{\bullet,\bullet}$ we are moving towards the
Rayleigh bound.  Eventually we will cross the line\footnote{This is a
straight line because $\beta - \gamma -1 = \sqrt{-\epsilon_{*}}$.}
$-(\beta-\gamma-1)^{2} = \epsilon_{*}$ and violate the  bound.
The same argument that was made in the P\"{o}schl-Teller model tells
us that there is a state $\ket{\beta_{*},\gamma_{*},\epsilon_{*}}$
such that
\begin{equation}
	a_{\beta_{*},\gamma_{*}} \ket{\beta_{*},\gamma_{*},\epsilon_{*}} 
	=0. 
	\label{eq:a-eq-0}
\end{equation}
Moreover, this state saturates the Rayleigh bound $\epsilon_{*} =
-(\beta_{*}-\gamma_{*}-1)^{2}$, \emph{i.e.}, $(\beta_{*},\gamma_{*})$
is on the boundary line.  The normalized solution to \eqref{eq:a-eq-0}
is
\begin{equation}
	\braket{x}{\beta_{*},\gamma_{*},\epsilon_{*}} = 
	\frac{1}{\mathcal{N}_{\beta_{*},\gamma_{*}}}\; 
	\frac{\sinh[\gamma_{*}+1](r)}{\cosh[\beta_{*}](r)}\quad\text{where}\quad
	\mathcal{N}_{\beta_{*},\gamma_{*}}^{2} = \int_{0}^{\infty} 
	\frac{\sinh[2\gamma_{*}+2](r)}{\cosh[2\beta_{*}](r)}\; dr\,.
	\label{eq:bound-sol}
\end{equation}
The asymptotic behavior of the function above as $r\to\infty$ is
$e^{-(\beta_{*}-\gamma_{*}-1)\,r} = e^{-\sqrt{-\epsilon_{*}}\,r}$
since the Rayleigh bound is saturated.  Also note that it behaves as
$r^{\gamma_{*}+1}$ as $r \downarrow 0$ as expected.  Beginning with
this state we can operate sequentially with
$a_{\bullet,\bullet}^{\dagger}$ to generated other states with the
same energy eigenvalues.  We are typically interested in states
$$a_{\beta_{*}+l,\gamma_{*}-l}^{\dagger}
a_{\beta_{*}+l-1,\gamma_{*}-l+1}^{\dagger}\dotsm
a_{\beta_{*}+2,\gamma_{*}-2}^{\dagger}
a_{\beta_{*}+1,\gamma_{*}-1}^{\dagger}
\ket{\beta_{*},\gamma_{*},\epsilon_{*}}.$$ This process is described 
in Figure~\ref{fig:angle}.

We can now make some observation about the three parameters $m$, $J$
and $\omega$ in the original equation \eqref{eq:our-gen-PT}, and the
allowed saturated bound values $\epsilon_{*}$, $\beta_{*}$ and
$\gamma_{*}$.
\begin{align}
	\sqrt{-\epsilon_{*}} &= m + \frac{q-1}{2}\,, \\
	\gamma_{*} &= J + \frac{q-4}{2}\,, 
	\label{eq:g-star}\\
	\beta_{*} &= \omega_{*} - \frac{1}{2} = m + J +q -\frac{3}{2}\,, 
	\intertext{where}
	\omega_{*} &= m + J + (q-1)\,.
	\label{eq:omega-J-m}
\end{align}
Equation~\eqref{eq:omega-J-m} is a rewriting of the saturated bound.
In particular note that for $q \ge 3$, the frequency $\omega_{*} \ge m
+ J +2 \ge 2$ and it is an integer.  Therefore we have that $\beta_{*}
\ge 3/2$ and $\beta_{*} \in \mathbb{Z} + \frac{1}{2}$.  The first
remark is that once $m$ is specified, $\epsilon_{*}$ is determined and
we have the Rayleigh bound line in Figure~\ref{fig:angle}.  The
various points on the bound line are associated with the different
values of $J=0,1,2,\dotsc$ according to eq.~\eqref{eq:g-star}.  Note
that action of $a_{\beta+1,\gamma-1}^{\dagger}$ takes $(\beta,\gamma)$
to $(\beta+1,\gamma-1)$ this has the consequences that $\omega \to
\omega +1$ and $J \to J-1$.  We know that $J \ge 0$ so the process
stops by at $\gamma_{\text{min}} = (q-4)/2$ because smaller $\gamma$
are outside the parameter range specified by the original hamiltonian
\eqref{eq:our-gen-PT}. 

	\section{More detail on solving the inhomogeneous equation for the first-order correction \label{sec:solvingInhom}}	
	The general solution to \eqref{eq:diffeqXi} depends on the type of potential chosen in the action. Here we discuss the procedure to solve this equation in our symmetry-breaking toy model, where $U(\phi) = \frac{\Lambda}{8}\qty(\phi^{2} - \phi_{0}^{2})^{2}$. When $q=1$, the ODE for the correction is \eqref{eq:odeXi_1-1}:
	\begin{equation*}
	\xi''(\nu)+\tanh(\nu) \xi'(\nu) + \frac{\tilde{\mu}}{\pi^{3}} \arctan(\sinh(\nu)) \qty(\pi^{2} - 4 \arctan(\sinh(\nu))^{2}) = 0.
	\end{equation*}
	We solved the equation by making use of a coordinate transformation. Using $x=\arctan(\sinh(\nu))$, the equation becomes
	\begin{equation*}
	\xi ''(x) + \frac{\tilde{\mu}}{\pi^{3}} x\sec[2](x) (\pi^{2} - 4x^{2}) = 0.
	\end{equation*}
	The boundary conditions for this equation are $\xi(x) \xrightarrow{\abs{x} \to \pi/2} 0$. Using Mathematica, a solution can be found to this equation. When translated back into the original coordinates, the expression is \eqref{eq:solnXi_1-1}. 
	
	In the case of $q=2$, the substitution is carried out with $x=\tanh(\nu)$ with the boundary conditions $\xi_{2}(x) \xrightarrow{\abs{x} \to 1} 0$.In this form the ODE is:
	\begin{equation}
	\tilde{\mu} \tanh(\nu) \sech[2](\nu) + 2\xi''(\nu) + 4\tanh(\nu) \xi'(\nu) = 0. \label{eq:odeXi_2-1}
	\end{equation}
	This is in a form that Mathematica can solve. It returns:
	\begin{equation}
	\xi_{2}(x) = \frac{1}{4} \tilde{\mu} ^2 \left(x \log \left(x^2-1\right)-x \log (4)-\log (1-x)+\log (x+1)+i \pi \right). \label{eq:solnXi_2-1-x}
	\end{equation}
	Transforming back to the original coordinates gives the solution in the text, \eqref{eq:solnXi_2-1}.
	
	\begin{figure}[tbp]
		\centering
		\includegraphics[width=\linewidth]{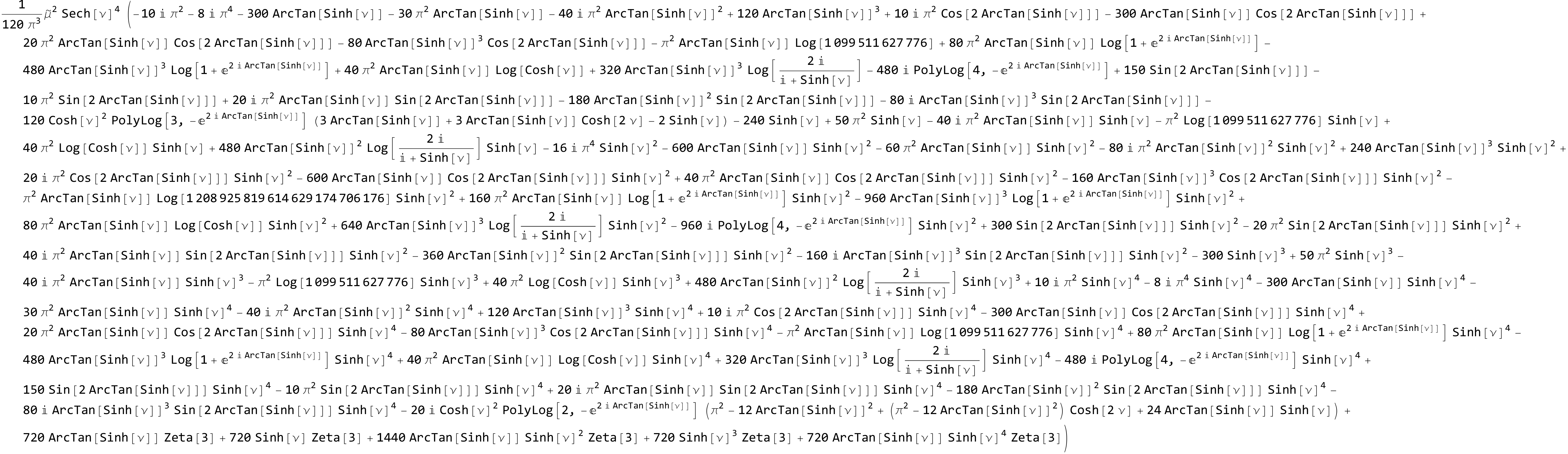}
		\caption{The solution that Mathematica returns for the case $q=3$. This function is real, but we did not find it to be illuminating.}
		\label{fig:xi-3-1-correction-nu-soln}
	\end{figure}
	
	For completeness, note that this same procedure can be carried out in other dimensions. We were able to find a solution to the case of $q=3$ by using the substitution $x=\sinh(\nu)$. The solution is lengthy, see figure~\ref{fig:xi-3-1-correction-nu-soln}.
	
	We did not find this solution illuminating, but we did verify that it was real, see figure~\ref{fig:xi-123-1-correction-nu}. Note that in some cases, Mathematica and other computer solvers can return a solution with an imaginary part. However, in all cases, this is merely a complex superposition of  solutions of the homogeneous equation  and so it can be subtracted off without loss of generality.

\providecommand{\href}[2]{#2}\begingroup\raggedright\endgroup
\end{document}